\documentclass[journal]{IEEEtran}
\usepackage{url}  
\usepackage{graphicx}  

\usepackage{soul}

\usepackage{color-edits}
\addauthor{pg}{red}
\addauthor{sc}{purple}
\addauthor{ac}{blue}
\addauthor{dh}{orange}
\addauthor{gq}{red}
\usepackage{multirow}
\usepackage{booktabs}

\usepackage{url}

\usepackage{colortbl}
\usepackage{subfig}

\definecolor{yellowgreen}{rgb}{0.67, 1, 0.18}
\definecolor{lightsalmon}{rgb}{1, 0.62, 0.47}
\definecolor{tomato}{rgb}{1, 0.38, 0.18}

\usepackage{tikz}
\def\checkmark{\tikz\fill[scale=0.4](0,.35) -- (.25,0) -- (1,.7) -- (.25,.15) -- cycle;}

\begin{document}

\title{ArduCode: Predictive Framework for Automation Engineering}
\author{
Arquimedes Canedo\textsuperscript{1}${^*}$~\thanks{${^*}$These authors contributed equally to this work.}~ 
Palash Goyal\textsuperscript{2}${^*}$~
Di Huang\textsuperscript{2}${^*}$~
Amit Pandey\textsuperscript{1}~
Gustavo Quiros\textsuperscript{1}~\\
\textsuperscript{1}{Siemens Corporate Technology}\\
\textsuperscript{2}{USC Information Sciences Institute}\\
}

\maketitle

\begin{abstract}
Automation engineering is the task of integrating, via software, various sensors, actuators, and controls to automate a real-world process. Today, automation engineering is supported by a suite of software tools including integrated development environments (IDE), hardware configurators, compilers, and runtimes. These tools focus on the automation code itself, but leave the automation engineer unassisted in their decision making. This can lead to \textcolor{black}{longer software development cycles due to} imperfections in the decision making \textcolor{black}{that arise when integrating} software and hardware. To address this \textcolor{black}{problem}, this paper \textcolor{black}{addresses} multiple challenges often faced in automation engineering and proposes \textcolor{black}{machine learning-based} solutions to assist engineers tackle these challenges. We show that machine learning can be leveraged to assist the automation engineer in classifying automation \textcolor{black}{code}, finding similar code snippets, and reasoning about the hardware selection of sensors and actuators. We validate our architecture on two real datasets consisting of 2,927 Arduino projects, and 683 Programmable Logic Controller (PLC) projects. Our results show that paragraph embedding techniques can be utilized to classify automation using code snippets with precision close to human annotation, giving an $F_1$-score of 72\%. Further, we show that such embedding techniques can help us find similar code snippets with high accuracy. Finally, we use autoencoder models for hardware recommendation and achieve a $p@3$ of 0.79 and $p@5$ of 0.95. \textcolor{black}{We also present the implementation of ArduCode in a proof of concept user interface integrated into an existing automation engineering system platform.}
\end{abstract}

\textit{\textbf{Note to practitioners---}} This paper is motivated by the use of artificial intelligence methods to improve the efficiency and quality of the automation engineering software development process. Our goal is to develop and integrate \textit{intelligent assistants} in existing automation engineering development tools to minimally disrupt existing workflows. Practitioners should be able to adapt our framework to other tools and data. Our contributions address important practical problems: (a) we address the lack of realistic datasets in automation engineering with two publicly available data sources; (b) we make the reference implementation of our algorithms publicly available on GitHub for other practitioners to have a starting point for future research; (c) we demonstrate the integration of our framework as an add-on to an existing automation engineering toolchain.

\section{Introduction}
Industrial automation is undergoing a technological revolution referred to as the \textit{fourth industrial revolution}~\cite{industriefourpointzero, hype}. The first revolution was mechanization of production enabled by steam and water power. The second revolution was the mass production enabled by electricity. The third revolution was automated production enabled by electronics and information technologies. The fourth revolution is smart production enabled by recent breakthroughs in intelligent robotics, sensors, big data, advanced materials, edge supercomputing, internet of things, cyber-physical systems, and artificial intelligence. These systems are currently being integrated by software into factories, power grids, transportation systems, buildings, homes, and consumer devices. 

\textit{Automation engineering software} (AES) integrates various sensors, actuators, and control with the purpose of automating real-world processes~\cite{AES1, AES2}. AES development has several challenges that distinguish it from general purpose software development. In this paper, we focus on two of the most prominent challenges: (a) AES software development is done by automation engineers, not by software experts; and (b) AES software interacts with the physical world by sampling sensors and writing outputs to actuators in well defined time intervals. These \textcolor{black}{challenges} have important implications in AES engineering. On the one hand, it is time consuming to develop automation code. This is partly because automation code is often not engineered for reusability. Thus, similar functionality is often developed from scratch. On the other hand, the interaction with the physical world requires automation engineers to understand the \textit{hardware configuration} that defines how sensors, actuators, and other hardware are connected to the digital and analog inputs and outputs of the system. This often requires an iterative engineering approach between the hardware and the software since any change in the hardware (e.g., a change of a component) has \textcolor{black}{an impact} in the software (e.g., input/output mappings), and vice-versa. The tight coupling between the hardware and the software produces longer development cycles in AES.

The lifecycle of industrial automation systems is divided into two phases: engineering and runtime. Engineering refers to all activities that occur \textit{before} the system is in operation. These engineering activities include hardware selection, hardware configuration, automation code development, testing, and simulation. Runtime, on the other hand, refers to all activities that occur \textit{during} the system's operation. These runtime activities include control, signal processing, monitoring, prognostics, etc. Applications of artificial intelligence (AI) in industrial automation have been \textcolor{black}{primarily} focused on the runtime phase due to the availability of large volumes of data from sensors. For example, time series forecasting algorithms have been very successful in signal processing~\cite{ARMA, ARMAMIT}. Planning and constraint satisfaction are used in controls and control code generation~\cite{Planning, Planning2}. Anomaly detection algorithms are becoming very popular in cyber-attack monitoring~\cite{anomaly2, anomaly1}.
Probabilistic graphical models and neural networks for prognostics and health management of complex cyber-physical systems~\cite{PGM1} have been deployed for systems such as wind~\cite{PHM2} and gas turbines~\cite{PHM1}.

The use of machine learning in the engineering phase, on the other hand, has remained relatively unexplored.
There may be several reasons for this.
First, engineering data is very scarce because of its proprietary nature~\cite{RESCOM}.
Second, the duration of the engineering phase is short compared to the runtime phase; some industrial automation systems are in operation for more than 30 years.
Therefore, the engineering phase is often considered less \textcolor{black}{critical} than the runtime phase. 
Third, acquiring human intent and knowledge is difficult.
Capturing engineering know-how in expert systems \textcolor{black}{is} time consuming and expensive.

This paper introduces the use of machine learning methods in AES to address three key tasks: code classification, semantic code search, and hardware recommendation. First, we demonstrate code classification on two real AES datasets. We learn representation of AES code via \textit{document embedding} methods, using different artifacts such as function calls, includes, comments, tags, and the code itself. Then we train classifiers on the code embeddings to categorize code projects. Our results show that our approach captures code structure and it is comparable to human annotation prediction performance. Second, using the resulting code embeddings, we demonstrate a semantic code search capability for AES code capable of finding \textcolor{black}{syntactic and structurally} equivalent fragments of code. Third, we develop a hardware recommendation system to auto-complete partial hardware configurations. Our results show a $3\times$ higher precision than the baselines. The original contributions of this paper are as follows:

\begin{itemize}
    \item The introduction of three AES tasks where AI has a big impact potential: code classification, semantic code search, and hardware recommendation.
    \item An unsupervised learning AES code embedding approach based on \textcolor{black}{natural language processing} suitable for code classification and semantic code search.
    \item The comparison of two hardware recommendation approaches using Bayesian Newtorks and Autoencoders.  
    \item The evaluation of our AI models in two real AES datasets consisting of 2,927 Arduino projects~\cite{arduino}, and 683 Programmable Logic Controller (PLC) programs~\cite{OSCAT}.
    \item The ArduCode reference implementation in Python\footnote{\url{https://github.com/arducode-aes/arducode}}, and datasets for advancing the AI research in automation consisting of: (i) AES source code and meta-data; (ii) an expert evaluation of code structural and syntactic similarity for 50 code snippets; (iii) a manually curated silver standard for hardware recommendation systems with two levels of granularity.
\end{itemize}

This paper is organized as follows. Section~\ref{sec:background} frames ArduCode in terms of \textcolor{black}{the state-of-the-art in AES, and} recent developments in \textit{code learning}. Section~\ref{sec:AES} gives an overview of the two types of AES systems: \textcolor{black}{industrial automation systems, and maker automation systems}. Section~\ref{sec:AESL} presents the proposed ArduCode architecture and methodology. \textcolor{black}{Section~\ref{sec:results} evaluates ArduCode's three AES learning tasks.} \textcolor{black}{Section~\ref{sec:cogeng} describes a proof-of-concept implementation of these tasks in an AES tool}. Section~\ref{sec:discussion} describes future directions of AI in automation engineering. Section~\ref{sec:conclusion} provides the concluding remarks.

\section{\textcolor{black}{Preliminaries and} Related Work}\label{sec:background}
To the best of our knowledge, we are the first to investigate the use of machine learning in AES. \textcolor{black}{However, there is a large body of work on AES. In this section we motivate the use of machines to assist AES tasks, and frame our contributions relative to the state-of-the-art in AES and related fields.}

Over the last few years, manufacturing is transforming itself from centralized mass production into a distributed lot size one production. One of a kind products, uniquely customized by customers, are being produced on demand. This shift is creating an unprecedented need for innovation in the engineering phase. Today, despite being short in duration (relative to the runtime phase), \textcolor{black}{we have estimated that} the engineering phase contributes with about 50\% of the total cost of automation. Thus, using AI in engineering phase is \textcolor{black}{an important technological development} for lowering the cost of production. In mass production the engineering phase is done upfront (at the beginning of the lifecycle) and only once for a particular product. In distributed lot size one production, engineering is done in parallel with the runtime as the production system must be adapted to satisfy all the variability associated to one of a kind products. While flexible machines and autonomous production systems can help realize lot size one production, the engineering phase will become intertwined with the runtime phase in the future.

\subsection{\textcolor{black}{Code Classification}}
As production demands change rapidly, there is a need to efficiently integrate new functionality into production. Today, AES engineers invest a significant amount of time creating functional libraries \textcolor{black}{to organize code according to its functionality. Publicly available examples of such libraries are the Arduino Library~\cite{ArduinoLibrary}, PLCOpen~\cite{PLCOpen}, and OSCAT~\cite{OSCAT}. On the other hand, the majority of automation code functions are not neatly organized in libraries. In these cases, automation engineers rely on cloning code by copying and pasting~\cite{CodeClones}. Cloning code solves an immediate problem because it allows the software development to make progress, but it creates a long-term maintenance problem because engineers quickly forget what a code function does. To address this problem, the authors in~\cite{CodeClones} present a text-based system to detect code clones for IEC 61131-3 programs commonly used in PLC programming. Although they broadly identify clone classification as one of the tasks in clone analysis, their system seems to be primarily focused on identifying code clones. In general, the lack code classification tools in the automation domain motivate our work. This paper introduces the use of automated code classification to reduce the effort of creating and maintaining functional code libraries. AI-driven code classification can be used organize code snippets according to their functionality. That is, code snippets can be automatically labeled according to \textit{what} they do -- e.g., signal processing, signal generation, robot motion control, or any organization-defined functionality. Code classification can be integrated in an engineering tool in such a way that as soon as a new function is released by an engineer, it is automatically classified into a category of a library. The engineer can be in the loop to confirm or correct the classification.}

\subsection{\textcolor{black}{Semantic Code Search}}
Frequent reconfigurations of the production system demand a much higher degree of automation code \textcolor{black}{assurance. This puts automation engineers under higher pressure to produce code that works as intended in much shorter engineering cycles. This problem can be broken down into two steps: (i) identifying potential errors, and (ii) coming up with an alternative solution to solve these errors. In the automation context, static analysis tools have been used to identify AES programming errors and defects. These defects are often referred to as code smells or technical debt indicators~\cite{TechnicalDebt}. For example, the authors in~\cite{IECStaticAnalysis} present a tool to detect issues in IEC 61131-3  programs. Their approach uses pattern-matching on program structures; control-flow and data-flow analysis; and call graph and pointer analysis. Similarly, the authors  in~\cite{Arcade} present Arcade.PLC, a framework for the verification and analysis of PLC code that combines model-checking and static analysis. Unfortunately, these tools are primarily focused on solving the first half of the problem. Therefore, this motivates the development of new approaches to assist automation engineers in identifying alternative implementations that can fix the problems found by the static analysis tools.}

Broadly, recent advances in \textit{code learning} \textcolor{black}{have shown that semantic code search is viable using machine learning. Code learning} can be divided into two categories~\cite{chen2019survey}: (i) language specific models, and (ii) language independent models. Language specific models use knowledge of the languages used in the code to generate low-dimensional representations. For example, \textit{code2vec}~\cite{alon2019code2vec} constructs abstract syntax tree from the code for Java language for the purpose of predicting a method's name from its content. It deconstructs the tree into several paths and learns code embedding by aggregating the representations of these paths. \textit{func2vec}~\cite{defreez2018path} uses control flow graphs to generate embeddings of functions in C language. They utilize such representations to detect function clones. Similarly, Deeprepair~\cite{white2019sorting} use a combination of \textit{word2vec} on tokens and recursive encoder on abstract syntax tree for Java token embedding. They use the representation to automatically repair programs with bugs. Several other works, such as DeepFix~\cite{deepfixit}, use language specific code learning to identify bugs and programming errors in codes. DeepTyper~\cite{deeptyper} uses recurrent neural networks to perform type inference in dynamically typed languages such as Javascript and Python. On the other hand, language independent models focus on syntactic representation learning. For example,~\cite{harer2018automated} utilize \textit{word2vec} directly on tokens from code to learn their representations. They show that their model can help predict software vulnerabilities. \cite{chen2018remarkable} utilize a similar approach for the task of automated program repair. The authors in~\cite{allamanis2015suggesting} introduce a syntactic model based on logbilinear contexts to generate new method names using these embeddings. Such models which do not use language syntax to learn code representations are less widely used compared to language specific models and often do not perform as well. However, in this paper, we show that our proposed language independent model achieves high accuracy in automation engineering tasks.

\subsection{\textcolor{black}{Hardware Recommendation}}
Automation engineering is the task of integrating various hardware components in software to achieve a production goal. \textcolor{black}{The selection of hardware components occurs early on in the engineering process~\cite{RESCOM}. There are two tasks associated to hardware configuration: (i) the selection of a specific hardware component (e.g., temperature sensor model A); (ii) the configuration of that hardware in terms of inputs and outputs for the automation software to interact with it. Depending on the complexity of the project, the selection of components may be done by the mechanical engineering department. However, the configuration of inputs and outputs is always done by the automation engineers. Also, note that the input and output configuration is tightly coupled the hardware selection. An automation program written for a given hardware is not guaranteed to work for another hardware selection. Therefore, any hardware configuration change triggers a re-engineering process~\cite{MechatronicsInconsistencies}.}

Today, AES engineers use an iterative process that is repeated several times because either the hardware was incomplete or the hardware selection was wrong. \textcolor{black}{Thus, reducing these hardware configuration iterations motivates the need for hardware recommendation systems. The goal of hardware recommendation systems is to predict a full hardware configuration from a partial hardware configuration. Recommendation or recommender systems are widely deployed in a variety of areas such as social media, video streaming, music streaming, news, dating, and consumer products~\cite{RecommenderSystems}. Unfortunately, recommender systems in the context of automation has remained a relatively unexplored area. The authors in~\cite{RESCOM} present RESCOM, a multi-relational recommender system for an industrial purchasing system. This system assists users in selecting the hardware for complex engineering solutions based on shopping basket statistical patterns and semantic information. The focus on purchasing makes RESCOM more suitable for solving the hardware selection sub-problem. To the best of our knowledge, we are the first to propose hardware recommender system for solving the hardware configuration for inputs and outputs sub-problem.}

\section{Automation Engineering Software Overview}\label{sec:AES}
There are two types of automation systems: industrial automation systems and \textit{maker} automation systems. This section introduces both. Despite some key differences, the most important aspect in common is the underlying computational model to interact with the physical environment. Most automation systems work on a periodic task model. Every period, or \textit{cycle}, is composed of three steps. The first step reads inputs from the hardware (e.g., rpm of a motor). The second step executes a \textit{task} every $T$ seconds. A task is similar to a thread and that executes a user-defined \textit{automation program} composed of one or more \textit{functions}. The third step writes outputs to the hardware (e.g., a control signal). These three steps realize the classic closed-loop control system configuration.

In addition to the computation and memory capabilities, an important measure to compare automation systems is the number of input/output (I/O) pins. The I/O capacity determines how many hardware devices can be wired to the automation system. I/O pins can be of type analog and digital. Automation systems can be interconnected to form industrial networks. Although communication can be done through the I/O pins, modern automation systems provide dedicated communication ports supporting Ethernet and industrial protocols such as Profinet~\cite{profinet} and OPC UA~\cite{opcua}.

AES programming is done through an integrated development environment (IDE) referred to as the \textit{engineering system}. This engineering system provides tools to support the AES development including programming language editors, hardware configurators, library managers, static analysis checkers, debuggers, compilers, and build systems. Improving the engineering systems with AI is the main focus of this paper.

\subsection{Industrial Automation Systems (IAS)}
These systems control industrial processes; many of these are safety-critical. IAS are real-time and must guarantee a response within a specified timing constraint (i.e., cycle). IAS are designed to operate for decades, without downtime, in very harsh environments of extreme temperatures, pressure, humidity, and vibration. Variations of these systems are developed by different manufacturers and are targeted to different industries. For example, discrete manufacturing use Programmable Logic Controllers (PLC), process plants use Distributed Control Systems (DCS), and power systems use Remote Terminal Units (RTU). 

Over the years, IAS have been programmed using a variety of programming languages. Most of these programming languages were conceived for automation engineers, not software engineers. Some of these languages, e.g. ladder logic, adopted domain-specific and graphical notations that were used to design relay racks in manufacturing and process control. Thus, by giving a syntax that automation engineers were familiar with, they were able to adopt these languages and write the AES code themselves. Today, these programming languages are standardized by the IEC61131-3 standard~\cite{iec611313}. Most vendors provide programming language interoperability and an automation program can contain functions written in different languages. This provides high flexibility to the automation engineers.

\subsection{Maker Automation Systems (MAS)}
The MAS market has been enabled by low cost electronics, microcontrollers, sensors, and actuators. \textit{Makers}, people who form a ``do it yourself'' (DIY) community and culture, have been the primary adopters of MAS. Arduino~\cite{arduino} is one of the most popular MAS. Its open source hardware and software is used by thousands of students and hobbyists around the world to develop DIY automation projects in robotics, home automation, entertainment, and wearables~\cite{banzi}.

MAS today are considered non-real time. However, there are a few recent advances to bring real-time operating systems (RTOS) to Arduino~\cite{freertosarduino, ARTE}. Despite this important difference w.r.t. IAS, MAS boards come with integrated analog and digital I/O pins, and follow the same computation paradigm used for automation systems. In the case of the Arduino IDE~\cite{arduinoIDE}, automation programs are referred to as \textit{sketches}. Sketches can be written in Arduino code files (INO), C, or C++. There is an extensive library of functions to work with hardware, manipulating data, and using control algorithms.

\textcolor{black}{MAS code is written by hobbyists with a very diverse backgrounds and skill sets. Comparing the quality of MAS and IAS code is not straightforward because they are subject to different programming environments. For example, Arduino's C/C++ environment provides the user full control over memory addressing whereas PLC environments constrain it. However, we noticed that code cloning~\cite{CodeClones} is a common pattern in the MAS community. The high availability of code for full projects makes it easy for hobbyists to copy and paste useful snippets. Many of these programmers give attribution to the original source through the use of comments in the code. A more in-depth analysis of MAS vs IAS code is an interesting future research direction.}

\section{ArduCode: Automation Engineering Software Learning}\label{sec:AESL}
\begin{figure*}[!htb]
\center{\includegraphics[width=0.9\textwidth]{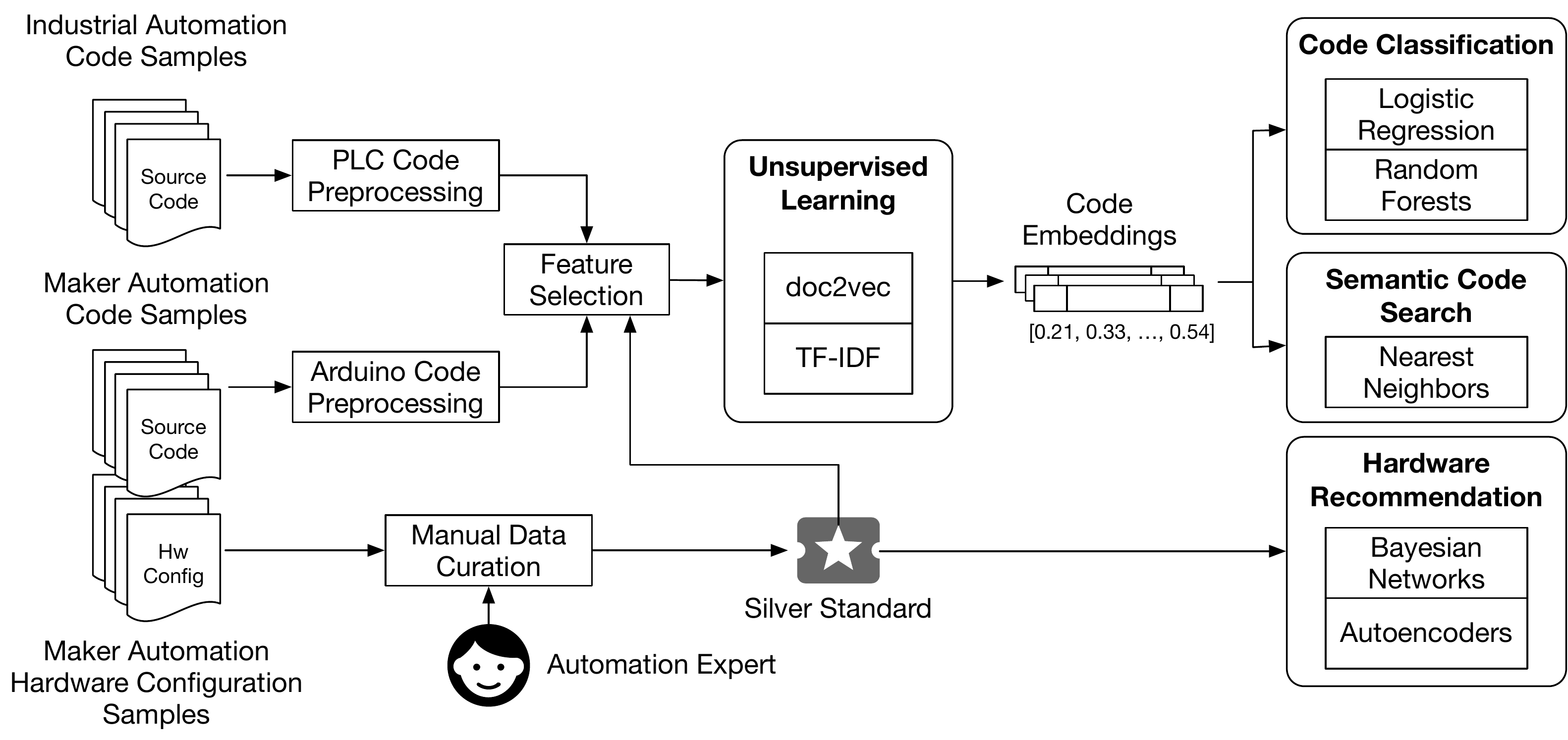}}
\caption{\label{fig:arducode_architecture}AES code learning architecture.}
\end{figure*}

In this section, we introduce our predictive framework for automation engineering (see Figure ~\ref{fig:arducode_architecture}). First, we provide our data collection methodology. Then, we describe our technical approach for each of the three automation engineering tasks: (i) code classification, (ii) semantic code search, and (iii) hardware recommendation.

\subsection{Data Collection}
To validate our approach, we collected two real datasets representative of MAS and IAS. The following subsections summarize the two datasets.

\subsubsection{Arduino Code}
\begin{figure}[!htb]
\center{\includegraphics[width=0.5\textwidth]{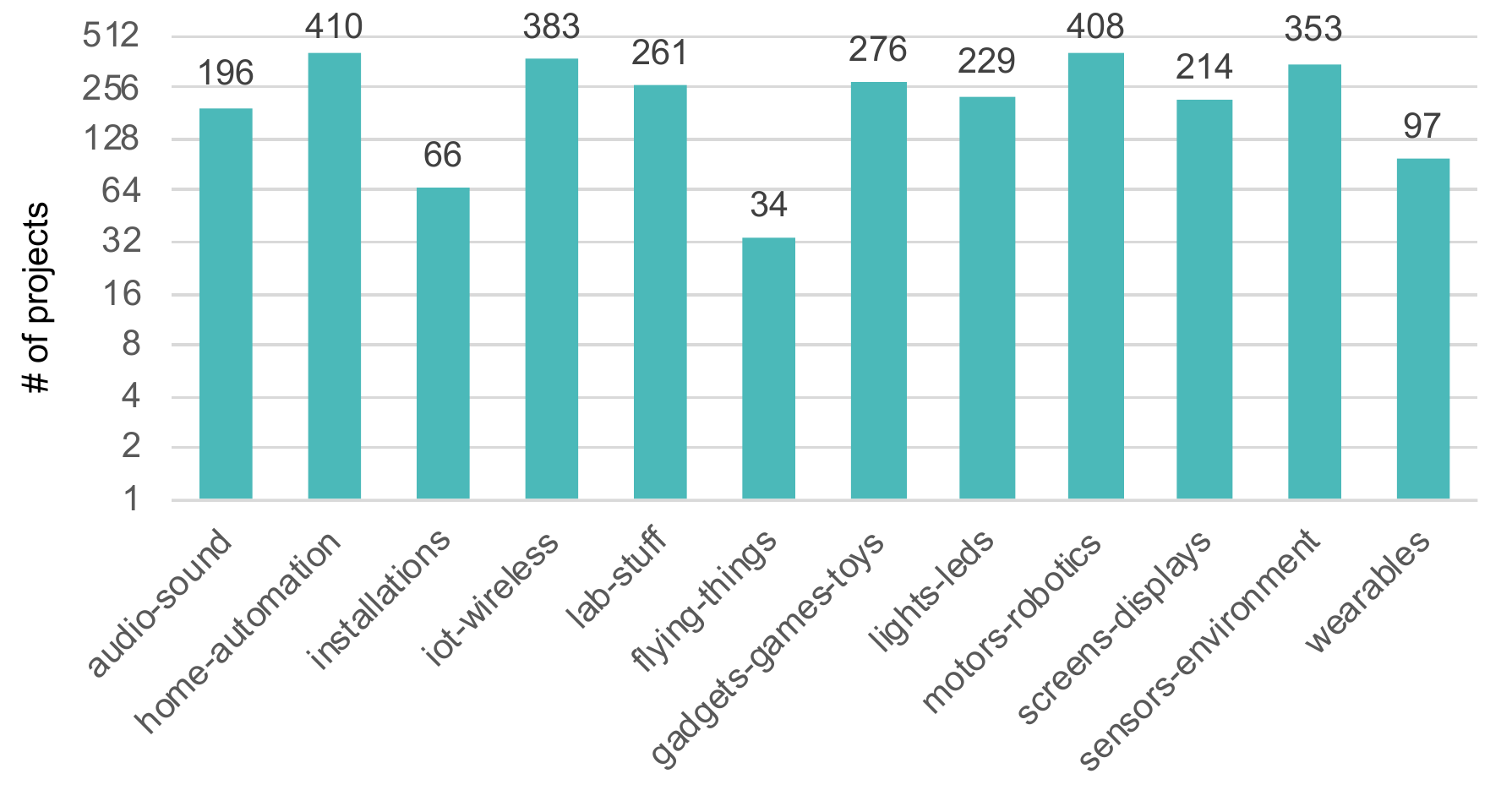}}
\caption{\label{fig:arduino_categories}Hand-curated Level-1 Arduino Project Categories.}
\end{figure}

We collected the source code and textual metadata from 2,927 Arduino projects from the Arduino Project Hub~\cite{ArduinoProjectHub}. The textual metadata consists of the project's category, title, abstract, tags, description, and hardware configuration. Each project is labeled by one category. In total, there are 12 categories as shown in Figure~\ref{fig:arduino_categories}. We use these categories as the labels to predict in the code classification task. Makers are well known for helping and fostering collaboration in the DIY community. The documentation associated to the Arduino projects is extensive. Therefore, the project's title, abstract, tags, and description metadata provide a upper bound baseline for label classification using human annotations.

The hardware configuration is a list of components required to build the project. In the 2,927 projects, there are 6,500 unique components. After manual inspection, we observed that different authors name the same component differently; e.g., ``resistor 10k'' and ``resistor 10k ohm''. To clean the data, we manually curated the hardware configuration lists and renamed the 6,500 components according to their functionality. \textcolor{black}{An important contribution of this paper is the definition of two} functional levels of abstraction for the hardware: \textcolor{black}{level-1, and level-2 functionality}. \textcolor{black}{Our} level-1 functionality consists of 9 categories: Actuators, Arduino, Communications, Electronics, Human Machine Interface, Materials, Memory, Power, Sensors. \textcolor{black}{Our} level-2 functionality further refines the level-1 into a total of 45 categories: Actuators.\{acoustic, air, flow, motor\}, Arduino.\{large, medium, other, small\}, Communications.\{ethernet, optical, radio, serial, wifi\}, Electronics.\{capacitor, diode, relay, resistor, transistor\}, Human Machine Interface.\{button, display, input, led\}, Materials.\{adapter, board, screw, solder, wiring\}, Memory.\{solid\}, Power.\{battery, regulator, shifter, supply, transformer\}, Sensors.\{accel, acoustic, camera, encoder, fluid, gps, misc, optical, photo, pv, rfid, temp\}. We use these two levels of hardware configuration to validate our hardware recommendation algorithm.

\subsubsection{PLC Code}
\begin{figure}[!htb]      
\center{\includegraphics[width=0.5\textwidth]{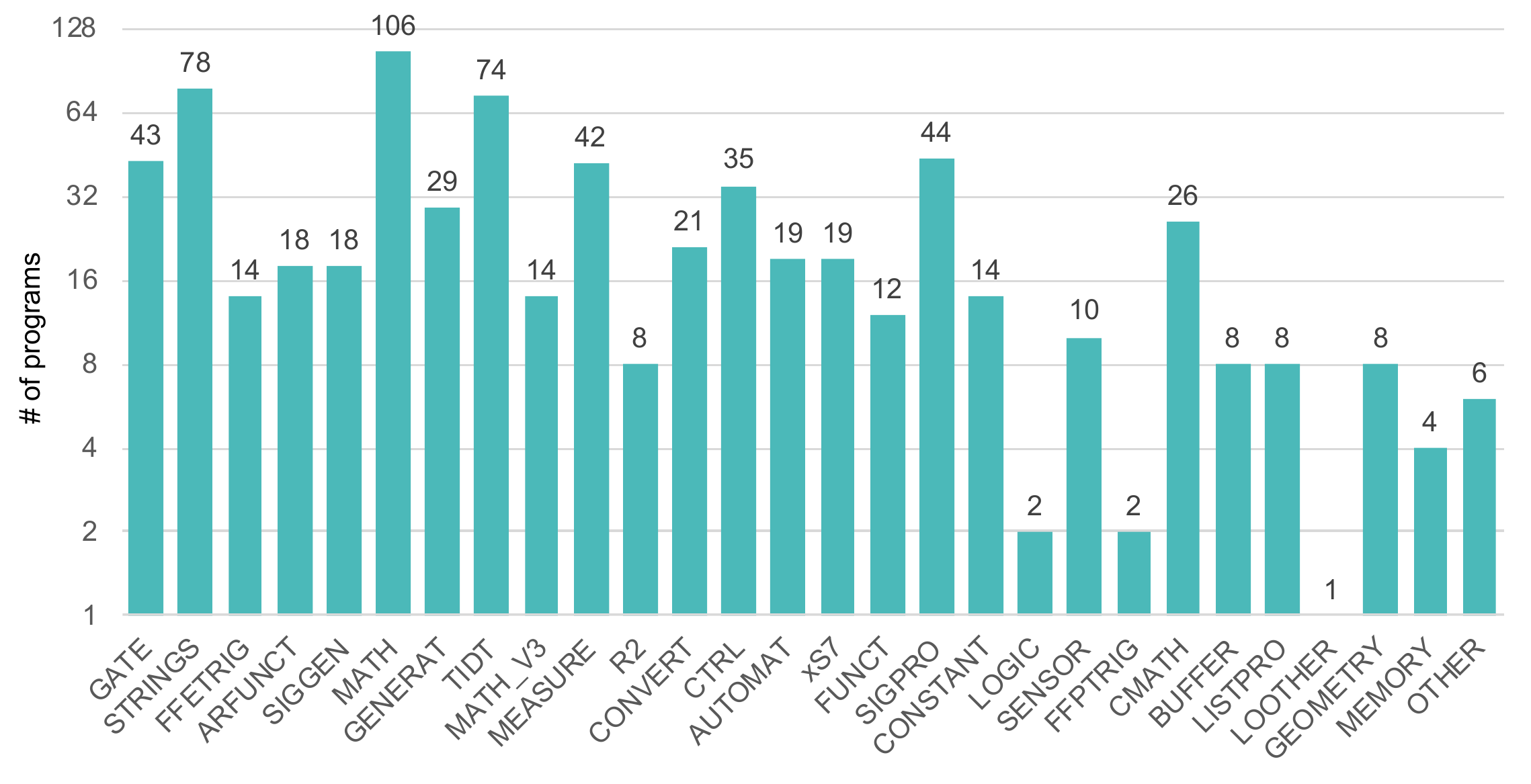}}
\caption{\label{fig:plc_categories}PLC function block categories.}
\end{figure}

The OSCAT library~\cite{OSCAT} is the largest publicly available library of PLC programs. The OSCAT-LIB is vendor independent and it provides reusable code functions in different categories such as signal processing (SIGPRO), geometry calculations (GEOMETRY), and string manipulation (STRINGS). These categories are extracted from the comment's section of each file marked by the line ``FAMILY: X'', where ``X'' is the category associated to that function. This line is eliminated from the dataset during training. In total, the OSCAT-LIB Basic version 3.21 contains 683 functions and 28 category labels. Figure~\ref{fig:plc_categories} shows the label distribution. 
The OSCAT-LIB does not contain hardware configuration, and therefore it is only suitable for the tasks of code classification and semantic code search. All the code is written in SCL language.

\subsection{Code Classification}
Given a code snippet, the task of code classification is to predict its label. \textcolor{black}{A label is a category associated to the code snippet such as the level-1 and level-2 Arduino project categories, or the OSCAT library function categories}. Our \textcolor{black}{machine learning} pipeline consists of four steps: preprocessing, feature selection, code embeddings, and classification. 

\subsubsection{Preprocessing}
We preprocess the automation projects and code snippets to expose the various features shown in Table~\ref{tab:features}. The Arduino dataset contains more features than the PLC dataset. Therefore, not all features are available in the PLC dataset. For example, PLC code does not have includes, and project data such as tags, title, descriptions, and components is not available.

\begin{table}[h!]
    \centering
    \caption{Features exposed by preprocessing.}
    \begin{tabular}{lccl}
        \toprule
        \textbf{Feature} & \textbf{Arduino} & \textbf{PLC} & \textbf{Description} \\
        \toprule
        Includes & \checkmark & -- & C/C++ includes\\
        Functions & \checkmark & \checkmark & Function names\\
        Comments & \checkmark & \checkmark & Comments in code\\
        Tokens & \checkmark & \checkmark & All code tokens\\
        Code & \checkmark & \checkmark & Code keywords\\
        LOC & \checkmark & \checkmark & Lines of code\\
        Tags & \checkmark & -- & Project tags\\
        Title & \checkmark & -- & Project title\\
        Descriptions & \checkmark & -- & Project descriptions\\
        Labels & \checkmark & \checkmark & Labels to predict\\
        Components & \checkmark & -- & Hardware configuration\\
        \toprule
    \end{tabular}
    \label{tab:features}
\end{table}

\subsubsection{Feature Selection}
The purpose of feature selection is to provide the ArduCode framework with a feature space exploration mechanism to compare the performance of different code representations in the task of code classification and semantic code search. The quality of the code embeddings is expected to vary according to the provided features. Therefore, the feature selection generates different experiments by combining different sets of features. For example, code can be represented by different combinations of includes, functions, comments, tokens, keywords. Code documentation can be represented by combinations of tags, titles, and descriptions. \textcolor{black}{Alternatively,} code representations and code documentation features can be combined \textcolor{black}{to generate richer feature vectors for the code embeddings}.

\subsubsection{Code Embeddings}
\textcolor{black}{In machine learning, the projection of a high-dimensional input into a lower-dimensional representation space is called an embedding.} The next step is to embed the textual representations generated by the feature selection \textcolor{black}{into a learned vector representation suitable for classification}. We compare the performance of the embeddings generated by gensim doc2vec ~\cite{le2014distributed} with the embeddings generated by the term frequency–inverse document frequency (tf-idf). \textcolor{black}{Doc2vec is a state-of-the-art method for distributed representations of sentences and documents -- i.e., tokens and code. Tf-idf, on the other hand, is a statistical measure that evaluates the relevance of a word is to a document -- i.e., tokens and code. Tf-idf is typically used as a baseline for validating other machine learning models. Both doc2vec and tf-idf generate an n-dimensional vector representation of the input.} The doc2vec's hyperparameters of interest are the embedding dimension, and the training algorithm (distributed memory and distributed bag of words). We run all our experiments with negative sampling of 5; \textcolor{black}{this draws 5 noise words to help the model differentiate the data from noise.}

\subsubsection{Classification}
The final step is to train a supervised model for code classification using the code embeddings as the input samples, and the code labels as the target values. We compare the performance of logistic regression and random forest classifiers using the $F_1$-score metric. \textcolor{black}{Logistic regression and random forest are two of the main workhorses for supervised learning classification. The $F_1$-score is the harmonic mean of the precision and recall. An $F_1$-score $=1$ represents perfect precision and recall.}

\subsection{Semantic Code Search}
Given a code snippet, the goal of semantic code search is to find similar programs. For automation engineering, similarity is defined in terms of syntax and \textcolor{black}{structure}. Syntax similarity helps engineers find useful functions in a given context, and \textcolor{black}{structural} similarity informs engineers on how other automation solutions have been engineered. 

\textcolor{black}{As shown in Figure~\ref{fig:arducode_architecture}, semantic code search uses the code embeddings generated by doc2vec to identify similar documents.} Doc2vec attempts to bring similar documents close to each other in the embedding space. For a given code embedding of a code snippet, the nearest neighbors are expected to be similar and \textcolor{black}{this distance metric is used as the} basis for our semantic code search. While this approach is intuitive for syntactically similar documents, it is unclear whether functional structure is captured in the embeddings.

\subsection{Hardware Recommendation}
Given a partial list of hardware components, the task of the hardware recommendation is to give a prediction of other hardware components typically used in combination with the partial list. The hand curated silver standard described above is used to learn the joint probability distribution of the hardware components. The hardware recommendation task is then to compute the conditional probability of missing hardware components given a partial list of components.

We compare two approaches for the hardware recommendation task. Our baseline consists of the predictions given on random hardware configurations. First, we learn a Bayesian network where the random variables are the hardware component categories. \textcolor{black}{Bayesian networks are a probabilistic graphical model that represent a set of variables (e.g., hardware components) and their conditional dependencies (e.g., 80\% chance that a sensor of type X is connected to an controller of type Y) via a directed acyclic graph.} We use the Pomegranate Python package~\cite{Pomegranate} to learn the structure of the Bayesian network and fit the model with 70\% of the hardware configuration data. The Bayesian network for level-1 components consists of 9 nodes and the one for level-2 components consist of 45 nodes. However, we were only able to fit the level-1 Bayesian network as the initialization of the Bayesian network takes exponential time with the number of variables. Typically, this cannot be done with more than two dozen variables due to the super-exponential time complexity with respect to the number of variables, and the level-2 hardware configuration consists of 45 variables. To overcome this limitation, we use an autoencoder implemented in Keras~\cite{Keras}. \textcolor{black}{An autoencoder neural network is an unsupervised learning algorithm that tries to learn a function to reconstruct its input. In our autoencoder,} the encoder learns a lower dimensional representation of the hardware configuration data, and the decoder learns to reconstruct the original input from the lower dimensional representation. \textcolor{black}{Overfitting is when a statistical model is tailored to a dataset and is unable to generalize to other datasets.} To avoid overfitting \textcolor{black}{in the autoencoder}, we use L1 and L2 regularizers.

\section{\textcolor{black}{Results}}\label{sec:results}
\textcolor{black}{This section evaluates the performance of the three AES tasks in ArduCode: code classification, semantic code search, and hardware recommendation. Our results establish the baselines for these tasks in the AES domain.}

\subsection{Code Classification}
\begin{figure}[!htb]
\center{\includegraphics[width=0.5\textwidth]{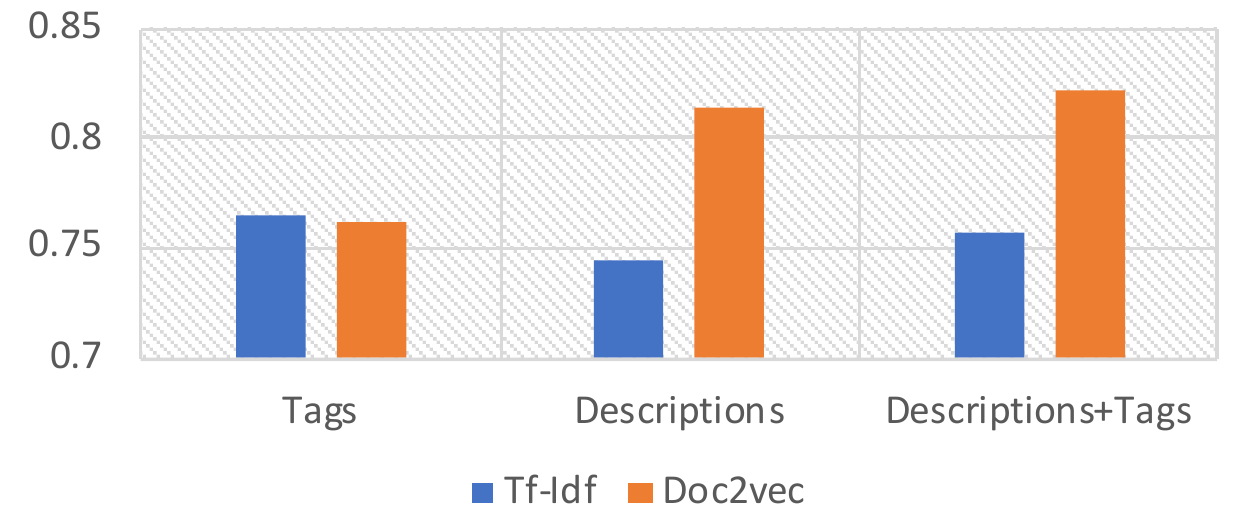}}
\caption{\label{fig:arducode_human}Human annotation prediction performance.}
\end{figure}

First, we established the lower and upper bounds for code label classification. The lower bound is given by training the code label classifier using random embeddings. The upper bound is given by training the code label classifier using human annotations. The Arduino dataset provides human annotations in the form of tags and descriptions that can be combined in three configurations: tags, descriptions, and descriptions+tags. We first embed these three configurations using tf-idf and doc2vec, and compare the label classification performance using the $F_1$-score. As shown in Figure~\ref{fig:arducode_human} doc2vec yields a better performance than tf-idf. The embedding dimension for doc2vec was set to 50, and the tf-idf models generated embedding dimensions of 1,469 for tags, 66,310 for descriptions, and 66,634 for descriptions+tags. Intuitively, the descriptions+tags configuration provides the upper bound $F_1$-score of 0.8213.

\begin{figure}[!htb]
\center{\includegraphics[width=0.5\textwidth]{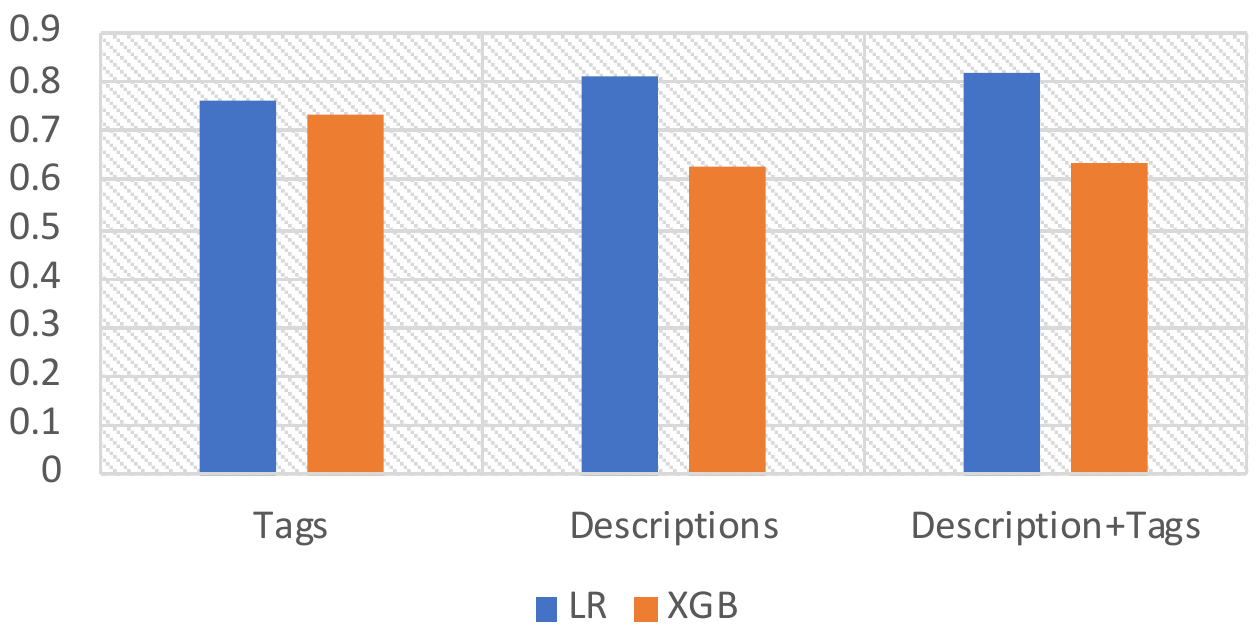}}
\caption{\label{fig:arducode_classifiers}Performance variation wrt classifier.}
\end{figure}

Figure~\ref{fig:arducode_classifiers} compares the performance of a logistic regression classifier (LR) from scikit-learn~\cite{scikitlearn} and a random forest classifier (XGB) from XGBoost~\cite{XGBoost} using the 50-dimensional doc2vec embeddings of tags, descriptions, and descriptions+tags. \textcolor{black}{Our implementation in these libraries was driven purely by the convenience of their software interface and their popularity in the machine learning community.} \textcolor{black}{An implementation using alternative libraries should yield similar results to the ones obtained in our implementation}. \textcolor{black}{Our results show that} the LR classifier provides significantly better performance than the XGB. 

\begin{figure}[!htb]
\center{\includegraphics[width=0.5\textwidth]{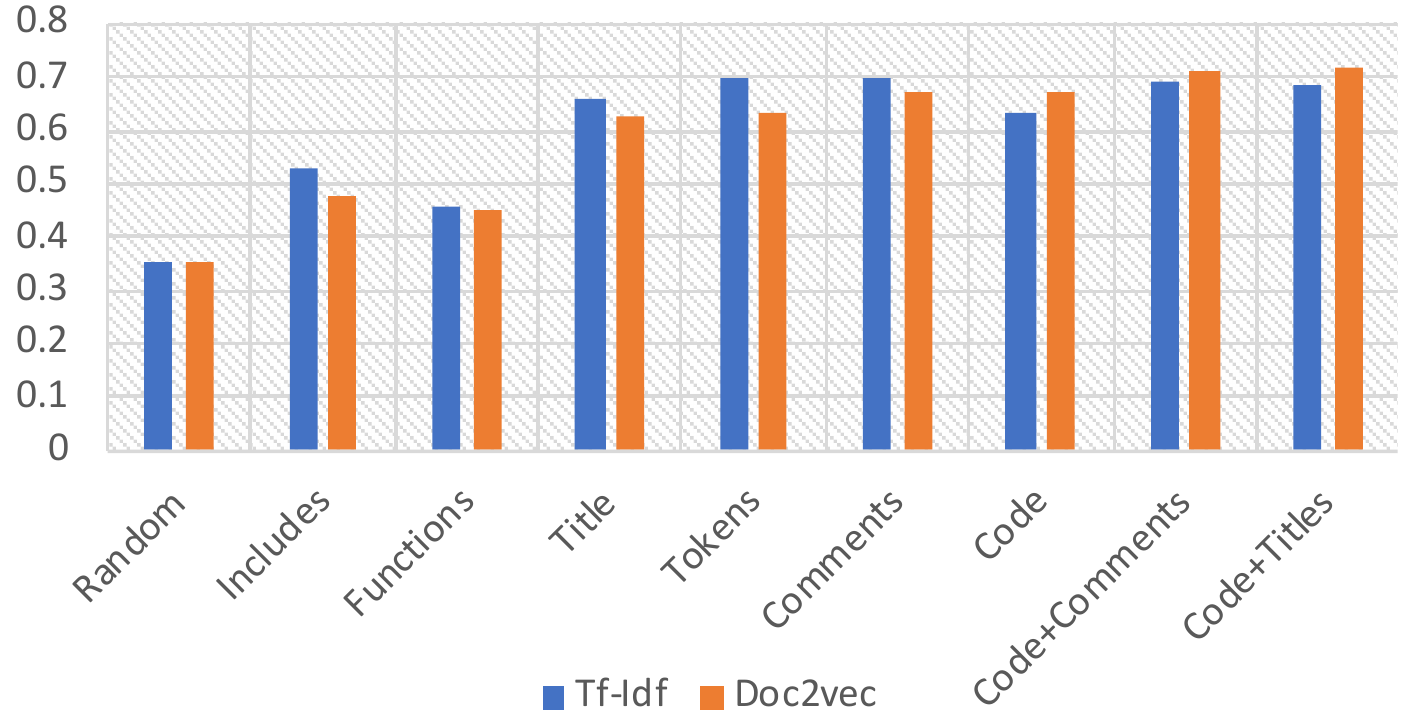}}
\caption{\label{fig:arducode_prediction}Code label prediction using code.}
\end{figure}

We establish the lower bound by generating 50-dimensional random embeddings and predicting the labels using the LR classifier. Figure~\ref{fig:arducode_prediction} shows that with both tf-idf and doc2vec the lower bound is 0.3538. After establishing the upper and lower bounds, we use different code features to predict labels. Figure~\ref{fig:arducode_prediction} shows that embedding includes and functions provide a slightly better performance than the random baseline due to the very limited amount of information contained in these: 1.82 includes and 4.70 functions on average. Other code features improve the classification accuracy significantly. For example, tokens and code are similar representations and give a similar $F_1$-score of 0.63 and 0.67. These results also show that comments contain valuable information that can be used to predict the code label with a score of 0.67. Embedding code+comments and code+titles yield the highest $F_1$-scores of 0.71. These results show that the prediction performance with code feature embeddings is comparable to human annotation embeddings with improvements of 2.03$\times$ and 2.32$\times$ over the random baseline, respectively.

\begin{figure}[!htb]
\center{\includegraphics[width=0.5\textwidth]{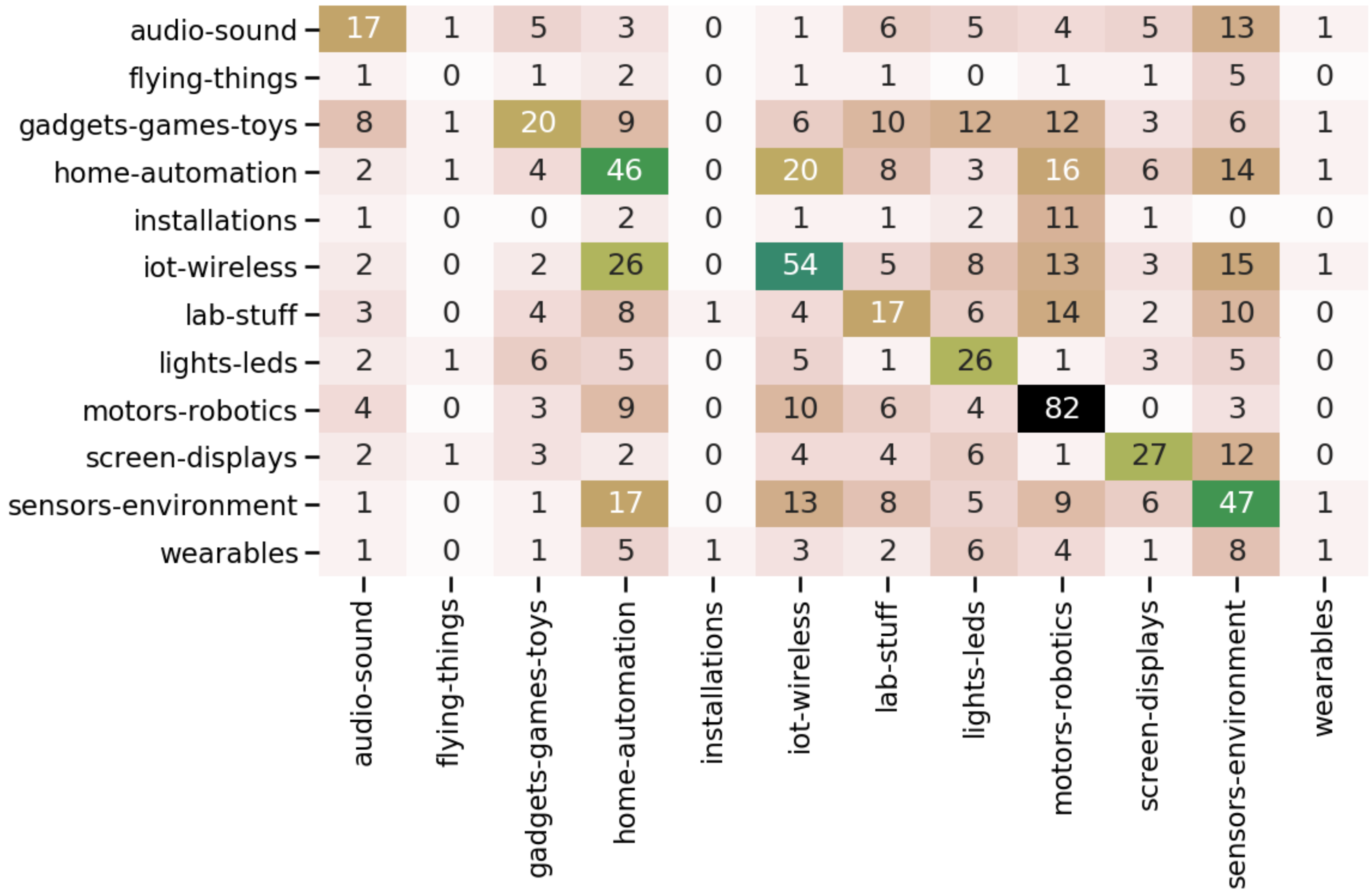}}
\caption{\label{fig:confusion}ArduCode classifier's confusion matrix.}
\end{figure}

The classifier's confusion matrix for code feature embeddings using doc2vec is shown in Figure~\ref{fig:confusion}. Although the matrix is diagonally dominant, the dataset is imbalanced in terms of number of samples per class as shown in Figure~\ref{fig:arduino_categories}. In particular, the classifier's poor performance in the categories flying-things, installations, and wearables correspond to their small number of samples in the dataset.

Since the PLC dataset does not have any human annotation features, we can only compare the performance of code feature embeddings (against the random baseline. The $F_1$-score for the code embeddings is 0.9024 and for the random baseline is 0.2878; a 3.13$\times$ improvement. Compared to the Arduino dataset, the PLC dataset has less samples (683 vs 2,927), more category labels (28 vs 12), and less lines of code per file on average (55 vs 177). These are three factors that influence the higher prediction accuracy of ArduCode on the PLC dataset.

\subsection{Semantic Code Search}
To validate the quality of our code embeddings, we randomly sampled 50 Arduino code snippets, and tasked a group of 6 software engineers to score the similarity of each code snippet to its top-3 nearest neighbors. For every code snippet pair, two similarity ratings for \textit{code syntax} and \textit{code structure} are given. A rating of $1$ represents similarity, and a rating of $0$ represents the lack of similarity. Code syntax refers to the use of similar variables and function names. Code structure refers to the use of similar code arrangements such as if-then-else and for loops. In addition, every rating has an associated \textit{expert's confidence} score from 1 (lowest confidence) to 5 (highest confidence) that represent the expert's self-assurance during the evaluation. During the expert evaluation we eliminated 5 out of 50 samples where one of the top-3 nearest neighbors was either an empty file, or it contained code in a different programming language. We found three code snippets written in Python and Javascript. 

Table~\ref{tab:avgscores} shows the average code syntax and code structure similarity scores given by experts. We only report the high confidence ratings (avg. confidence $>= 4.5$) in order to eliminate the influence of uncertain answers. We also measure the experts' agreement via the Fleiss Kappa. These results show that the similarity scores for both syntax and structure are high for the top-1 neighbors (0.68 and 0.61 respectively) but reduce significantly (under 0.50) for the top-2 and top-3 neighbors. The experts are in substantial agreement ($0.61 <= \kappa <= 0.80$) in their syntax similarity scores, and in moderate agreement ($0.41 <= \kappa <= 0.60$ in their structure similarity scores. While these results confirm that doc2vec code embeddings capture syntactic similarity, they also show that some structure similarity is captured in the top-1 neighbor. \textcolor{black}{After their individual assessment, the experts gathered as a group to discuss their findings. Something that quickly became clear is that the experts' background contributed to the deviation in the ratings. Half of the experts with an automation background had additional insights that made them more confident and congruent in their structural similarity assessments. On the other hand, the other half of the experts without an automation background were less confident and congruent in their ratings. With additional context from the automation domain, two of the three of the non-automation experts expressed that this information would have made their assessment more confident and congruent.}

\begin{table}[h!]
    \centering
    \caption{Average code structure and code syntax similarity and Fleiss Kappa values for high confidence raters.}
    \begin{tabular}{llll}
        \toprule
          & \multicolumn{3}{c}{\textbf{Nearest neighbors}} \\
         \cmidrule{2-4} 
        \textbf{Similarity} & \textbf{Top 1 ($\kappa$)} & \textbf{Top 2 ($\kappa$)} & \textbf{Top 3 ($\kappa$)} \\
        \toprule
        Syntax & 0.61 (0.75) & 0.48 (0.70) & 0.32 (0.61) \\
        Structure & 0.68 (0.53) & 0.40 (0.44) & 0.33 (0.66) \\
        \toprule
        \end{tabular}
    \label{tab:avgscores}
\end{table} 

To further gain insight into our experiment, we selected four similar and three not similar code snippets, and measured the cosine similarity of their embeddings as shown in Table~\ref{tab:cosinedistance}. The selected code snippets have a strong agreement among the experts, and a high confidence in the similarity and lack of similarity across the top-3 nearest neighbors. These results confirm that the code snippets considered very similar by the experts are close to each other in the embedding space. On the other hand, code snippets considered not similar are far apart in the embedding space. 

\begin{table}[h!]
    \centering
    \caption{Code embedding cosine similarity for similar and not similar code snippets.}
    \begin{tabular}{lllll}
        \toprule
         & & \multicolumn{3}{c}{\textbf{Nearest neighbors}}\\
         \cmidrule{3-5} 
         &  & \textbf{Top 1} & \textbf{Top 2} & \textbf{Top 3}\\
        \toprule
        \multicolumn{5}{l}{\textbf{Similar code snippets}}\\
         & \#2696 & 0.8768 & 0.7527 & 0.7642 \\
         & \#547 & 0.8719 & 0.8705 & 0.8506 \\
         & \#2815 & 0.9465 & 0.9445 & 0.9126 \\
         & \#54 & 0.8513 & 0.8056 & 0.7815\\
        \toprule
        \multicolumn{5}{l}{\textbf{Not Similar code snippets}}\\
         & \#4512 & 0.5967 & 0.5497 & 0.5643 \\
         & \#4345 & 0.5415 & 0.4175 & 0.5192 \\
         & \#1730 & 0.5970 & 0.5035 & 0.5511 \\
        \toprule
    \end{tabular}
    \label{tab:cosinedistance}
\end{table}

\begin{figure}[!htb]
\center{\includegraphics[width=0.5\textwidth]{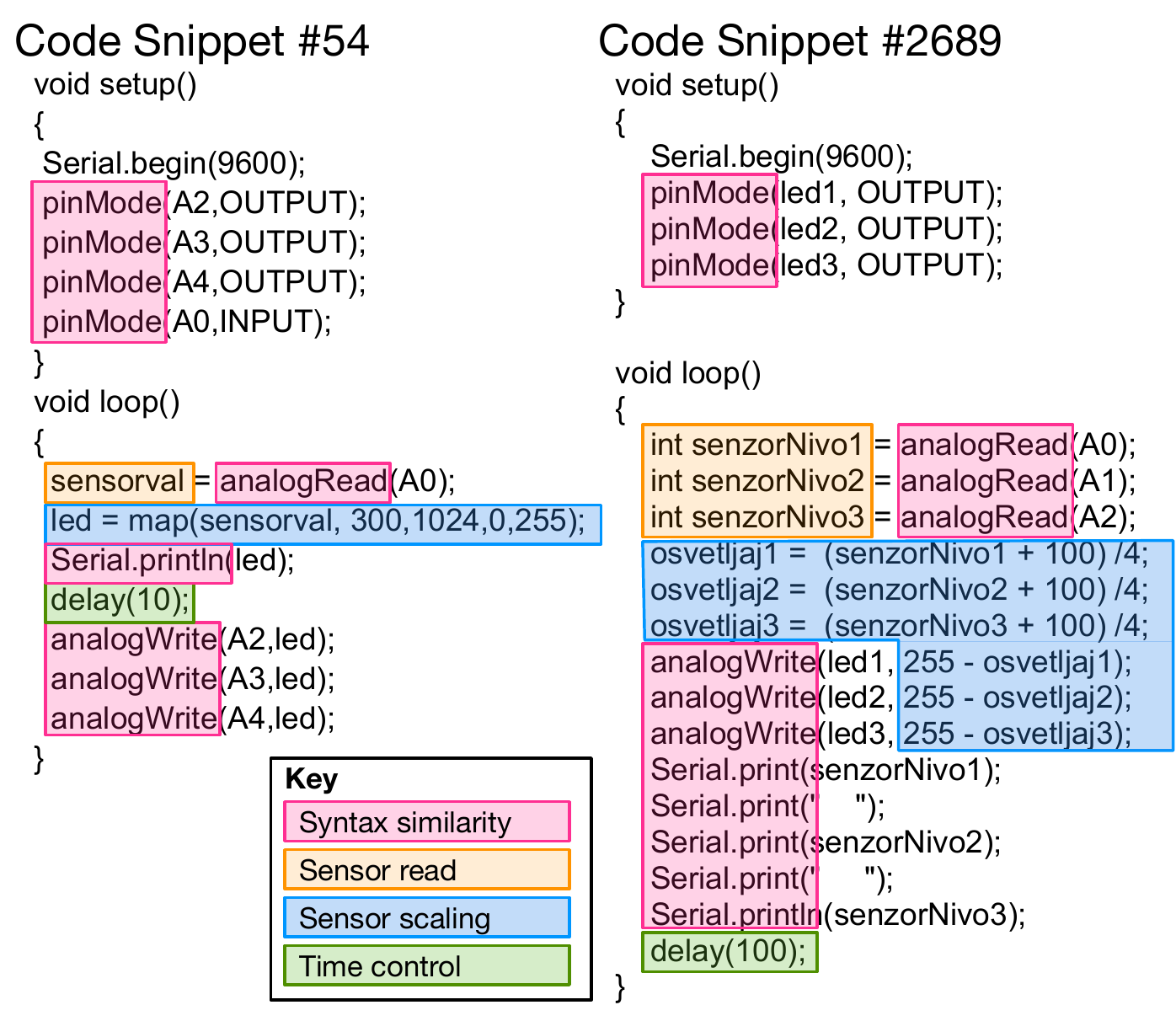}}
\caption{\label{fig:code_similarity_a}Arduino semantic code search result.}
\end{figure}

Figure~\ref{fig:code_similarity_a} shows two similar Arduino code snippets (\#54 and \#2689) produced by ArduCode. There are similarities between these two programs at different levels. First, all Arduino programs are required to have the \texttt{setup()} and \texttt{loop()} functions to initialize the program, and to specify the control logic executed on every cycle. Syntactically, the two programs use the same standard functions: \texttt{pinMode()} to configure the Arduino board pins where the hardware connects as inputs or outputs; \texttt{analogRead()} to read an analog value from a pin; \texttt{Serial.print()} to print an ASCII character via the serial port; \texttt{delay()} to pause the program for the amount of time (in ms) specified by the parameter; and \texttt{analogWrite()} to write an analog value to a pin. Semantically, the two programs read sensor values (only 1 value in \#54 and 3 values in \#2689), scale the sensor value to a range (from 300-1024 to 0-255 using \texttt{map()} in \#54 and to $(x+100)/4$ in \#2689), print the scaled sensor value via the serial port, write the analog value to a LED (a single LED in \#54 and three LEDs in \#2689), and pause the program (10ms in \#54 and 100ms in \#2689). Note that the order in which these operations are scheduled is different in the two programs. Functionally, the two programs perform the same task of creating a heatmap for a sensor value using LEDs. While there are some syntactic similarities, or semantic code search is also able to capture semantic and \textcolor{black}{structural} similarities.

\subsection{Hardware Recommendation}
In hardware recommendation, we are interested in recommending the top-k hardware components. Therefore, we evaluate our models in terms of $precision@k$. $Precision@k$ is the portion of recommended hardware components in the top-k set that are relevant. For each hardware configuration in the test data, we leave one hardware component out, and measure its $precision@k$. Table~\ref{tab:p_at_k} shows the results for the random baseline, the Bayesian network, and the autoencoder. As expected, the performance of the random baseline improves linearly from $p@1=0.1$, $p@3=0.32$, and $p@5=0.54$ to $p@9=1$ for the level-1 hardware predictions. The Bayesian Network also improves linearly from $p@1=0.32$, $p@3=0.59$, and $p@5=0.79$. The autoencoder provides both the best performance and the best improvements from $p@1=0.36$, $p@3=0.79$, and $p@5=0.95$. Note that the autoencoder's $p@3$ is the same performance as the Bayesian Network's $p@5$, $0.79$. Furthermore, the autoencoder achieves $>0.95$ precision at $p@5$, and the Bayesian network at $p@8$. 


\begin{table}[h!]
    \centering
    \caption{$p@k$ results for level-1 hardware predictions.}
    \begin{tabular}{llll}
        \toprule
        \textbf{$p@k$} & \textbf{Random} & \textbf{Bayesian} & \textbf{Autoencoder} \\
         & \textbf{Baseline} & \textbf{Network} & \\
        \toprule
        $p@1$ & 0.10 & 0.32 & 0.36 \\
        $p@3$ & 0.32 & 0.59 & 0.79 \\
        $p@5$ & 0.54 & 0.79 & 0.95 \\
        $p@9$ & 1.00 & 1.00 & 1.00 \\
        \toprule
    \end{tabular}
    \label{tab:p_at_k}
\end{table}

Learning a Bayesian Network for level-2 hardware components is computationally unfeasible. Therefore, we rely on an autoencoder to accomplish this task and the $p@k$ results are reported in Table~\ref{tab:autoencoder}. The overall $p@k$ performance of the autoencoder for level-2 is comparatively lower than for level-1. The reason is that the level-2 hardware configuration is sparser than level-1. On average, level-2 configurations have 4/40 components and level-1 configurations have 4/12 components. However, the improvement over the random baseline is of $10\times$ for $p@1$, $5\times$ for $p@3$, $4\times$ for $p@5$, and $3\times$ for $p@10$.

\begin{table}[h!]
    \centering
    \caption{$p@k$ results for level-2 hardware predictions.}
    \begin{tabular}{lll}
        \toprule
        \textbf{$p@k$} & \textbf{Random Baseline} & \textbf{Autoencoder} \\
        \toprule
        $p@1$ & 0.02 & 0.21 \\
        $p@3$ & 0.06 & 0.34 \\
        $p@5$ & 0.11 & 0.45 \\
        $p@10$ & 0.21 & 0.69 \\
        \toprule
    \end{tabular}
    \label{tab:autoencoder}
\end{table}

\section{\textcolor{black}{Proof of Concept Implementation - Cognitive Automation Engineering System}}\label{sec:cogeng}

\begin{figure*}[!ht]
\center{\includegraphics[width=1\textwidth]{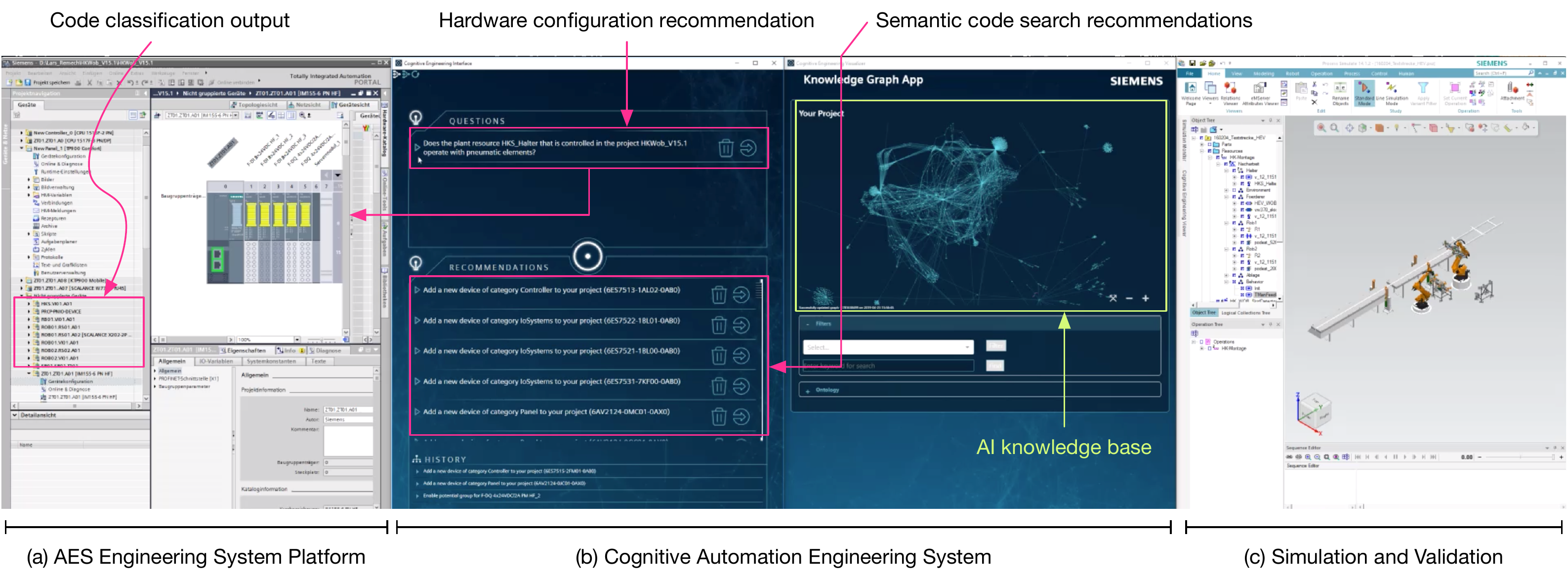}}
\caption{\label{fig:cogeng}Cognitive Automation Engineering System.}
\end{figure*}

This section presents our experience in developing a proof of concept system that attempts to transition this research into an application. Our goal is to explore concrete implementation ideas to assist the AES engineer during the development process in such a way that the AI-driven methods do not disrupt the original AES workflows. 

Figure~\ref{fig:cogeng} shows our implementation. 
The system is centered around an engineering knowledge base, which is built in two complementary ways: by analyzing past engineering projects, and by direct input and curation of engineering know-how by experts.
\textcolor{black}{We leverage semantic web technologies and represent the engineering knowledge in an RDF triple store.
Engineers with sufficient expertise can edit the graph directly, and we aim to develop interfaces that simplify the formalization of engineering knowledge by domain experts.}
Note that the past engineering projects and expert know-how are specific to the organization where the system would be deployed and therefore the ArduCode's models would reflect that prior expertise.
Leveraging data from past projects is a very powerful approach because it allows the integration of existing libraries and unstructured repositories of functions into ArduCode's learning process. Thus, code classification, semantic code search, and hardware recommendation can be continuously improved over time as more data becomes available from the regular engineering process. On the other hand, the direct expert input allows the resolution of issues when ArduCode's confidence is low in a prediction, and the direct expert curation can enable the improvement of ArduCode's models over time.

We use an existing AES engineering system as a platform (see Figure~\ref{fig:cogeng}(a)). This approach has multiple advantages. First, an existing AES Engineering System provides well known graphical user interfaces and workflows that engineers are familiar with. Second, we can leverage the APIs and open interfaces from the AES Engineering System to push and pull data from the current and other automation projects. Pulling data is critical for building a robust knowledge base for the AI methods to learn from. Pushing data is equally critical because it allows the AI methods to assist the engineer within the well established workflows and interfaces.  Third, it provides a development environment for plug-ins where novel concepts for AI-assistant user interfaces can be developed and tested with real users.

The proof of concept user interface plug-in that we developed in shown in Figure~\ref{fig:cogeng}(b). The goal of this user interface is to provide the AES engineer an \textit{assistant} that is continuously monitoring the state of the AES project and providing recommendations from code classification, semantic code search, and hardware recommendation. 
\textcolor{black}{The assistant system reads the user's automation project containing hardware configuration and program code. Multiple analysis components are then executed on the data and generate recommendations based on this available information.}
As the user edits configurations or automation code, the code classifier, semantic code search and hardware recommendation are invoked, and the user interface collects the results of the analyses and presents them to the user in an organized manner based on the context of the corresponding engineering task.
In this user interface, the main interaction occurs through questions and recommendations. The AI-assistant presents questions to the user, which the user can answer providing new information to the system.
\textcolor{black}{Questions to the user are triggered by information known to be missing in the knowledge model. For instance, if an analysis depends on the value of an attribute of the project (e.g. safety integrity level), and this attribute is missing in the current project, then the assistant will issue a question to the user asking for the value of this attribute (e.g. What is the safety integrity level of this project?).}
Based on the information gathered from the engineering tool and from user answers, the assistant generates recommendations that the user either accepts or rejects. 
\textcolor{black}{When the user provides the project’s SIL level as an answer, the system may trigger additional analyses using this information (e.g. determine hardware to recommend for the required safety level, classify the code in the automation project to separate the safety and non-safety code and reorganize the project structure accordingly, or find library code suitable for the safety application and recommend this code to the user).}
When the user accepts a recommendation, an action is taken and the project is modified accordingly. When the user rejects or ignores a recommendation the system simply continues its operation. Note that these interactions are also useful for refining the AI models. These interactions can be used in the future as data for a lifelong learning system that self-improves over time. To keep the user informed on what the AI-methods do, we provide an interface to the AI knowledge base in a graphical form. We use this view to highlight the nodes and edges related to a given recommendation.

Just like in the traditional AES development process, validation through simulation plays an important role. We show that the simulation and validation loop can be tightly coupled to the proposed Cognitive Automation Engineering System. After a recommendation is accepted, and the change is pushed to the AES Engineering System, this can be validated in simulation. Figure~\ref{fig:cogeng}(c) shows a production line simulation that the AES project controls. This implementation shows that AI-assistants can be non-disruptive to the existing AES development process. 

\textcolor{black}{The assistant system has been used for empirical evaluation of the approach in a lab setting. Our plan is to evaluate this tool in a real production environment.}

\section{Discussion}\label{sec:discussion}
ArduCode is the starting point for predictive automation engineering tasks. This section discusses its known limitations and motivates several directions for future work.

\subsubsection{Capturing code structure} While doc2vec captures some structural code similarity as confirmed by our set of experts, there are other recent approaches such as code2vec that are likely to better capture code structure and improve the code classification and semantic code search tasks. However, parsing C++ code requires full access to all the library dependencies. In the case of most Arduino programs, resolving the include paths requires a major manual effort. In the future, we expect to develop some automation to generate the abstract syntax trees and embed the code using models such as code2vec. 

\subsubsection{Hardware-software gap}
ArduCode's hardware recommendation is limited to hardware components. This task would be even more useful if it incorporated software elements such as library or API recommendations. We obtained poor results with a supervised learning approach using hardware configuration as the input samples, and includes and function names as the target values. Without context such as descriptions or titles, this is a hard task even for experts because the software references hardware only through non-descriptive variables (e.g., A0, A1). To bridge this gap, one promising idea is to model software elements as random variables in the Bayesian Network and use expert know-how to define their conditional probabilities. \textcolor{black}{Another potential direction is to take into consideration wiring diagrams that describe how the hardware components are connected to the controller, and based on this information determine the connection to the software. However, these diagrams are often not synchronized to the software view and could introduce many inconsistencies. A second challenge is that these diagrams are often created using ad-hoc methods by the software developers and their syntax and semantics are not consistent.}

\subsubsection{\textcolor{black}{Continuous integration of new knowledge}}
\textcolor{black}{In the current architecture, ArduCode's model is updated via re-training. This means that new knowledge is not automatically integrated into ArduCode's knowledge base. The new knowledge is manually integrated through a re-training step. Re-training takes time and requires a machine learning expert. However, with the high-availability of cloud computing and graphic processing units (GPU), we do not anticipate re-training being a bottleneck for the automation domain. An interesting direction for future work is to extend ArduCode into a lifelong learning architecture~\cite{lifelonglearning}. Lifelong learning is the ability of a machine learning system to sequentially retain learned knowledge and to transfer that knowledge over time when learning new tasks and improve its capabilities. Such approach would continuously integrate new knowledge as it becomes available, instead of being limited to discrete re-training steps.}

\section{Conclusion}\label{sec:conclusion}
In this paper, we introduced and studied three automation engineering predictive tasks. First, we showed that our code classification approach based on doc2vec code embeddings and logistic regression achieves an $F_1$-scores of 72\% and 90\% on two real datasets. Second, a group of 6 experts validated the semantic code search task by assessing the syntax and structure similarity of 50 code snippets. Third, we demonstrated a $p@3$ of 79\% and $p@5$ of 95\% for the hardware recommendation task using an autoencoder. \textcolor{black}{Additionally, we implemented these tasks in a proof-of-concept implementation of a cognitive automation system. This system has been used for empirical evaluation of ArduCode in a laboratory setting.}

Future research directions are as follows. Evaluate ArduCode's doc2vec approach against recent approaches such as code2vec that are likely to better capture code structure and improve the code classification and semantic code search tasks. In addition, ArduCode's hardware recommendation is limited to hardware components. This task would be even more useful if it incorporated software elements such as library or API recommendations. One promising idea is to model software elements as random variables in the Bayesian Network and use data mining techniques on existing projects to define their conditional probabilities.

\section*{Acknowledgements}
We thank Evan Patterson, Jade Master, and Georg Muenzel for their valuable input and discussions.


\bibliographystyle{ieeetr}
\bibliography{bibliography}
\end{document}